\documentclass[
  preprint,
  amsmath,amssymb,
  aps,physrev,
  floatfix
]{revtex4-2}

\usepackage[english]{babel}
\usepackage{graphicx}
\usepackage{dcolumn}
\usepackage{bm}
\usepackage{hyperref}
\usepackage{setspace}

\begin{document}
\preprint{APS/123-QED}
\title{\textbf{Circulators Based on Coupled Quantum Anomalous Hall Insulators and Resonators} 
}%

\author{Luis A. Martinez$^{1,}$\footnote{Electronic mail: martinez289@llnl.gov}}
\author{Nick Du$^{1}$}
\author{Nicholas Materise$^{1}$}
\author{Sean O'Kelley$^{1}$}
\author{Xian Wu$^{1}$}
\author{Gang Qiu$^{2,3}$}
\author{Kang L. Wang$^2$}
\author{Gianpaolo P. Carosi$^1$}
\author{Tony Low$^3$}
\author{Dong-Xia Qu$^{1,}$\footnote{Electronic mail: qu2@llnl.gov}}

\affiliation{$^1$Lawrence Livermore National Laboratory, Livermore, CA 94550, USA}
\affiliation{$^2$Department of Electrical and Computer Engineering, University of California, Los Angeles, CA 90095, USA}
\affiliation{$^3$Department of Electrical and Computer Engineering, University of Minnesota, Minneapolis, MN 55455, USA}

\date{\today}

\begin{abstract}
Integrated plasmonics is advancing rapidly, enabling a wide range of functionalities to be incorporated onto a single chip. Applications span information processing, computation, quantum sensing, and dark-matter detection. This progress has driven the development of integrated non-reciprocal devices, which are essential for preventing unwanted feedback that can degrade system performance. While non-reciprocal devices have been realized in edge magnetoplasmon materials via classical interference effects, their operation is often limited by the input power range. Here, we demonstrate that topological circulators utilizing asymmetric coupling offer improved input power range, isolation, and insertion loss. In this configuration, we demonstrate that the coupling between a chiral edge magnetoplasmonic resonator and a pair of LC resonators is well described by an effective non-Hermitian two-site Hatano--Nelson model with asymmetric directional couplings, resulting in nonreciprocal behavior. The coherent photon-plasmon interaction enables a circulator with up to $50\,\mathrm{dB}$ of isolation across a broad range of excitation power. These results suggest that magnetic topological insulators provide a promising platform for realizing asymmetric non-Hermitian couplings at radio frequencies and for exploring regimes of strong directional suppression and possible exceptional-point physics. More broadly, they highlight the potential of topological-material-based microwave devices for future integration with superconducting quantum information platforms. 

\begin{description}
\item[Keywords]
Non-reciprocal devices, magnetic topological insulator, plasmons

\end{description}
\end{abstract}

\maketitle


\section{\label{sec:level1}Introduction}
The chiral nature of the edge magnetoplasmon makes it a promising candidate for advancing chip-scale non-reciprocal devices at quantum-classical interfaces \cite{Mahoney2017PRX, Mahoney2017, PhysRevResearch.6.013081}. Edge magnetoplasmon (EMP) propagation has been extensively studied in quantum Hall (QH) systems \cite{Ashoori1992, Kumada2011, Kumada2014}. Their coupling to external devices in the non-Hermitian regime has been investigated, while the possibility of accessing strongly asymmetric transport and exceptional-point-related phenomena remains largely unexplored. Recent studies have demonstrated that non-Hermitian systems, particularly those of the Hatano–Nelson type, offer potential for realizing isolators and switches because their intrinsically asymmetric coupling strengths can induce exceptional points (EPs), leading to highly non-reciprocal transport phenomena \cite{PhysRevLett.134.153801}. In this context, EMPs can serve as non-reciprocal links that asymmetrically couple radio-frequency (RF) resonators, enabling circulation or isolation behavior.


\begin{figure}[h]
\centering
\includegraphics[width=0.75\textwidth]{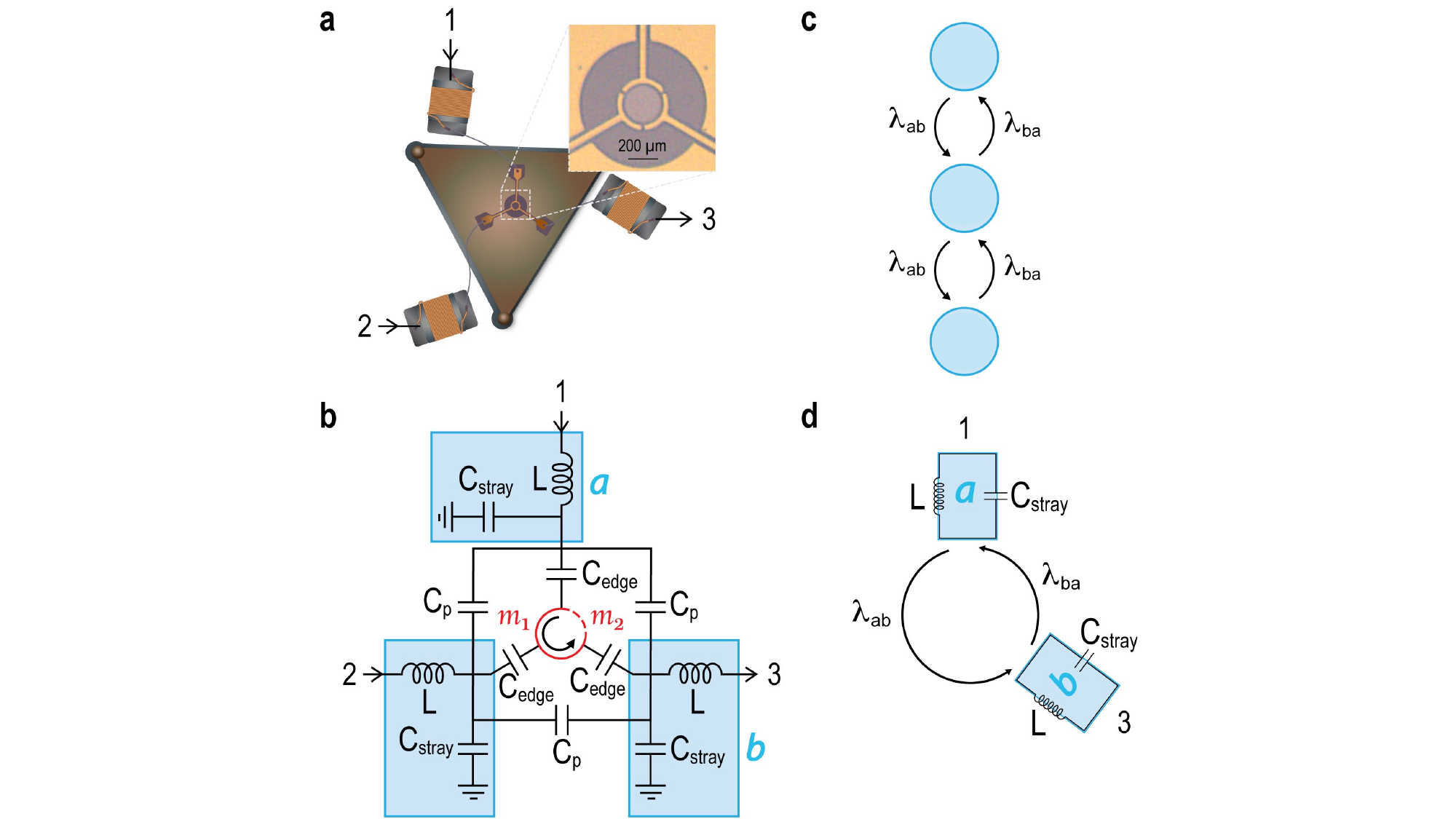}
\caption{{\bf Device and Hatano-Nelson model.} {\bf a,} Schematic of the three-port circulator device, consisting of a QAH mesa with an edge conductance of $e^2/h$, and three $180$ nH lumped-element wire-loop inductors are wire-bonded between the mesa and the transmission line of the printed circuit board. Edge magnetoplasmon resonant modes are excited and propagate along the edge of the circular mesa. The inset shows the optical image of the circulator, with a $100$-$\mu \mathrm{m}$-radius QAH mesa separated from the gold pad that defines the three signal ports by a $15~\mu m$ gap. {\bf b,} Equivalent circuit diagrams illustrating parasitic capacitance $C_{\mathrm{p}}$, stray capacitance $C_{\mathrm{stray}}$, and edge capacitance $C_{\mathrm{edge}}$. Components $a$ and $b$ (blue pads) represent the LC resonators at ports 1 and 3, respectively. $m_1$ and $m_2$ represent the anticlockwise transport of the EMPs from resonator $a$ to $b$, and from $b$ to $a$, respectively.  {\bf c,} Schematic of a one-dimensional (1D) Hatano-Nelson chain, where asymmetric couplings $\lambda_{ab}$ and $\lambda_{ba}$ between cells give rise to non-reciprocal transport in a non-Hermitian system. {\bf d,} Sketch of the circulator configuration composed of two LC resonators, $a$ and $b$, coupled non-reciprocally through the interference of EMP and the parasitic capacitive pathways, leading to asymmetric coupling $\lambda_{ab}$ and $\lambda_{ba}$.}\label{fig1}
\end{figure}

Compared to quantum Hall (QH) systems, quantum anomalous Hall (QAH) materials offer distinct advantages, such as maintaining a stable number of edge channels across a broad range of applied magnetic fields, including zero magnetic field \cite{doi:10.1126/science.1234414,RevModPhys.95.011002}. Moreover, QAH states can be engineered in multilayer magnetic topological insulators or heterostructures, providing design flexibility that can be tailored to meet specific circuit requirements \cite{osti_1783152, osti_10401313}. This makes them well-suited for integration with superconducting circuits, where magnetic fields can be detrimental to performance, and semiconductor devices, which can tolerate or even require moderate magnetic fields for operation. The QAH insulator, Cr-doped (Bi$_x$Sb$_{1-x}$)$_2$Te$_3$, provides a promising platform for realizing strong plasmon-photon interactions. This is due to its high-impedance plasmon mode, with an impedance of $Z = h/Ce^2 \sim 25.8 \ \rm{k} \Omega$ and a Chern number $C = 1$. Because the zero-point voltage fluctuation scales as $\sqrt{Z}$, this large $Z$ markedly amplifies the electric-dipole coupling between the plasmon and nearby quantum emitters, making high-impedance edge magnetoplasmonic resonators ideal for achieving strong asymmetric plasmon-photon coupling and opening up exciting opportunities in non-reciprocal electronics.

In this work, we investigate a coupled plasmon-photon system, in which an on-chip QAH plasmon resonator is asymmetrically coupled to a pair of LC resonators. While Mahoney \textit{et al}. studied a similar architecture in a QH system, they did not explore the connection to Hatano-Nelson-type couplings \cite{Mahoney2017PRX}. In our hybrid system, we observe a pronounced non-reciprocal transmission dip accompanied by a positive $\pi$ phase shift. By leveraging the chiral nature of the EMP modes, we interpret the observed behavior as being consistent with an effective two-site Hatano--Nelson--type plasmonic system \cite{PhysRevLett.77.570,PhysRevB.58.8384,PhysRevLett.134.153801}, in which asymmetric coupling may drive the system towards a strongly nonreciprocal non-Hermitian transport regime. Furthermore, the system achieves an \textit{average isolation} of $20\,\mathrm{dB}$ over an approximately $50\,\mathrm{MHz}$ bandwidth and a peak isolation exceeding $50\,\mathrm{dB}$ at $543.8$ MHz for an input microwave power of $-119\,\mathrm{dBm}$. The circulator remains operational down to an input power of $-149\,\mathrm{dBm}$. These results suggest a new mechanism for harnessing the intrinsic unidirectionality of QAH edge states to engineer non-reciprocal devices, with promising implications for scaling superconducting quantum processors and improving axion haloscope sensitivity.

\section{\label{sec:level1}Results}

\subsection{\label{sec:A}Circulator Configuration and Effective Hatano-Nelson-Type Coupling}

The configuration of the three-port circulator device is shown in Fig.~\ref{fig1}a, which consists of an EMP resonator with a radius of 100 $\mu$m and three lumped element inductors, each with an inductance $L=180$ nH. The inductor and the stray capacitor $C_{\mathrm{stray}}$ (Fig.~\ref{fig1}b) form an LC resonator, whose resonance frequency $\omega_{LC} = 1/\sqrt{LC_{\mathrm{stray}}}$ is set to closely match the EMP resonance frequency $\omega_{m}=\left[\frac{\sigma_{xy}}
{2\pi\epsilon_0\epsilon_{eff}} \left(\mathrm{ln}\frac{2}{|q|w}+1\right) + v_D\right]q$, where $q$ is the wave vector, $\sigma_{xy}=C e^2/h$ the Hall conductivity, $\epsilon_{eff}$ the effective dielectric constant, $v_D$ the drift velocity of the 1D edge state, and $w$ the physical width of the EMP mode at the edge. The value of $\omega_{m}$ can be directly extracted from experimental characterization of the circulator without requiring an impedance-matching circuit. At each port of the circulator, the EMP resonator is capacitively coupled to an LC resonator through an edge capacitor $C_{\mathrm{edge}}$, which is formed between the Au pad and the EMP resonator (Fig.~\ref{fig1}b). The microwave transmission coefficients $S_{31}$ and $S_{32}$ are characterized using a measurement scheme described in Appendices~\ref{app:A} and~\ref{app:B}. All cryogenic measurements were performed in a dilution refrigerator at a temperature of approximately $10\,\mathrm{mK}$.

The circulator architecture, consisting of two LC resonators ($a$ and $b$) and the two directional EMP channels, is described by a minimal non-Hermitian coupled-mode matrix that assumes idealized non-reciprocity propagation \cite{Lau2018, YYWang2021, PhysRevLett.134.153801, 2014NatPh..10..394P}:
\begin{equation}
 H_{\rm{eff}}/\hbar=
  \left( {\begin{array}{cccc}
   \omega_a-i\kappa_a & g & 0 & 0 \\
   0 & \omega_{m_1}-i\kappa_{m_1} & g & 0 \\
   0 & 0 & \omega_b-i\kappa_b & g \\
   g & 0 & 0 & \omega_{m_2}-i\kappa_{m_2} \\   
  \end{array} } \right)\label{eq1}
\end{equation}
\smallskip

\noindent
where $\kappa_{a}$ and $\kappa_{b}$ are the decay rates of the LC resonators $a$ and $b$, respectively. $\kappa_{m_1}$ and $\kappa_{m_2}$ are the decay rates of the EMP resonators along two propagation paths, $m_1$ and $m_2$, as illustrated in Fig. \ref{fig1}b. $\omega_{a, b, m_1, m_2}$ are the bare resonance frequencies of the LC and EMP modes, respectively, and $g$ denotes the coupling strength between the LC resonators and the EMP modes. In the frequency-domain coupled-mode treatment (see Appendix~\ref{app:E}), the response coefficients are written as $\alpha_j(\omega)= -i\omega + i\omega_j + \kappa_j = \kappa_j - i(\omega-\omega_j)$.

By eliminating the intermediate EMP modes, Eq.~\eqref{eq1} can be reduced to an effective two-site Hatano--Nelson-type model (Fig.~\ref{fig1}c) with unequal direction-dependent couplings between neighboring unit cells (see Appendix~\ref{app:E}), 
\begin{equation}
\lambda_{ab}(\omega)=\frac{g^2}{\alpha_{m_1}}, \qquad
\lambda_{ba}(\omega)=\frac{g^2}{\alpha_{m_2}},
\end{equation}
where $\alpha_{m_1}(\omega)$ and $\alpha_{m_2}(\omega)$ describe the frequency-domain complex response of the two EMP propagation paths. Because these paths have different lengths and dissipation, $\alpha_{m_1}\neq \alpha_{m_2}$ in general, leading to asymmetric effective coupling. In our effective model, the non-reciprocal inter-resonator couplings, $\lambda_{ab}$ and $\lambda_{ba}$, arise from destructive and constructive interference between the EMP and parasitic capacitive pathways (Figs.~\ref{fig1} b and d). As a result, they are determined by the EMP propagation lengths and depend sensitively on both the input microwave power and the device operating temperature. Within this effective model, sufficiently strong suppression of the one directional coupling could in principle drive the reduced two-site system towards an EP, marked by eigenvalue and eigenvector coalescence \cite{heiss2012physics}.

\subsection{\label{sec:B}Circulator Characterization}

\begin{figure}[h]
\centering
\includegraphics[width=0.9\textwidth]{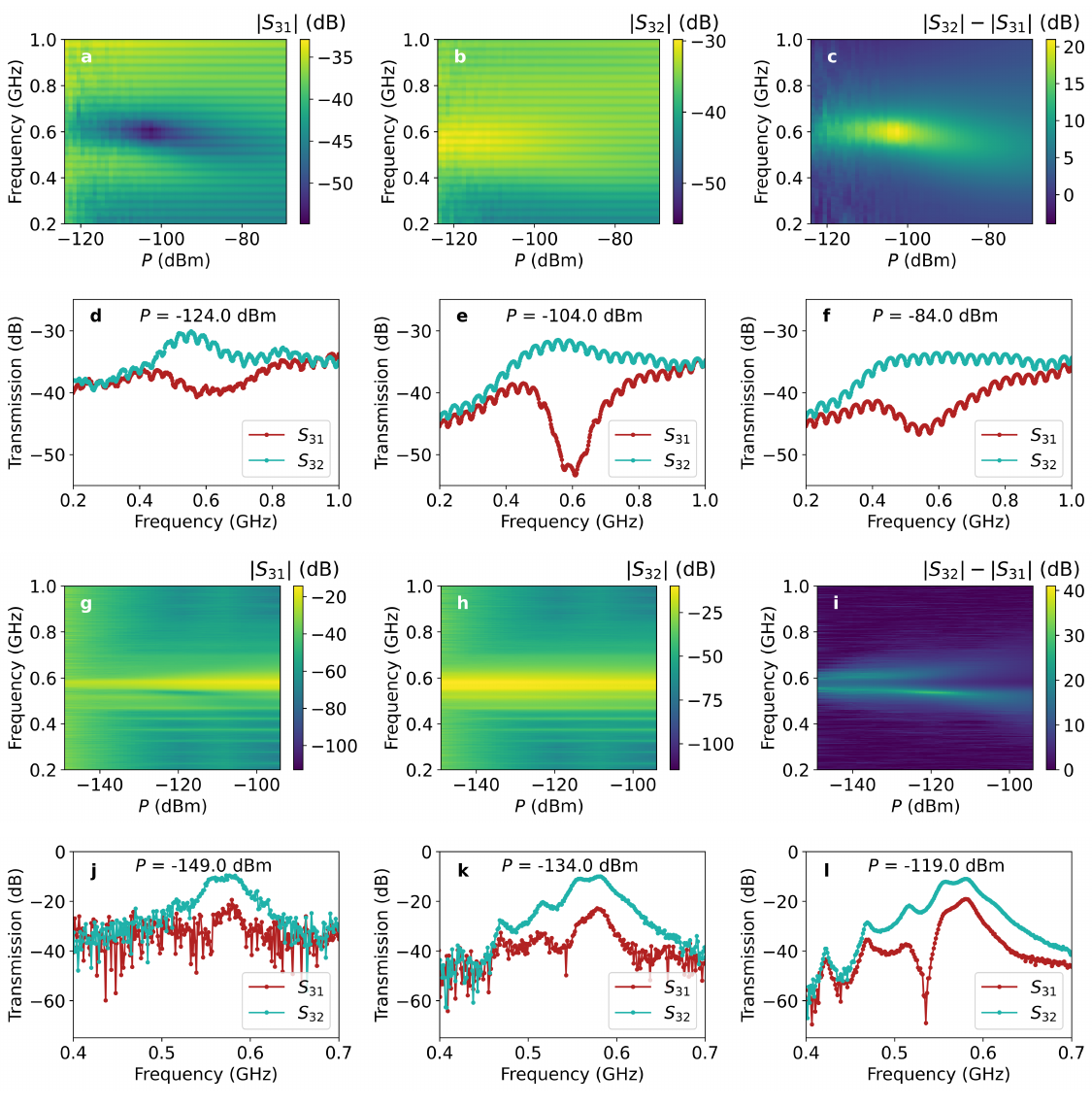}
\caption{{\bf Transmission response for circulators without and with coupling to LC resonators.} {\bf a--c,} Power-dependent microwave transmission responses $|S_{31}|$, $|S_{32}|$, and $|S_{32}|-|S_{31}|$ for a $100\,\mathrm\mu m$-radius circulator without coupling to LC resonators. The signal is excited from ports 1 and 2, and measured at port 3. {\bf d--f,} Frequency line cuts from panels {\bf a} and {\bf b}, shown for three representative input power levels. {\bf g--i,} Power-dependent microwave transmission responses $|S_{31}|$, $|S_{32}|$, and $|S_{32}|-|S_{31}|$ for a $100\,\mathrm\mu m$-radius circulator coupled to LC resonators. {\bf j--l,} Frequency line cuts from panels {\bf g} and {\bf h}, shown for three representative input power levels.}\label{fig2}
\end{figure}

We begin by examining the transmission in an EMP resonator, both without and with LC resonators, across a wide spectral range. Figures~\ref{fig2} a--c show the transmission through the EMP resonator, without coupling to an LC resonator. Owing to the constructive and destructive interference between the EMP mode (red traces), coupled via $C_{\mathrm{edge}}$, and the port-to-port RF signal transmitted through the parasitic capacitance $C_p$ (Fig.~\ref{fig1}b), $S_{31}$ and $S_{32}$ exhibit a non-reciprocal response with a bandwidth of approximately $100\,\mathrm{MHz}$ (Figs.~\ref{fig2} d--f). The isolation magnitude, $|S_{32}|-|S_{31}|$, reaches a maximum of $20\,\mathrm{dB}$ over an input-power range spanning approximately $5\,\mathrm{dBm}$. 

Figures~\ref{fig2} g--i show the transmission spectra as a function of frequency and input power when the EMP and LC resonators are coupled. Notably, the magnitude of $S_{32}$ remains nearly constant over an input power range spanning more than $50\,\mathrm{dBm}$. The circulator exhibits $\sim 20\,\mathrm{dB}$ of non-reciprocity over an input power range from $-150$ to $-90\,\mathrm{dBm}$, with an approximate bandwidth of $50\,\mathrm{MHz}$ (Fig.~\ref{fig2}i). This level of isolation and power handling is particularly important for quantum information science, where protecting qubits from amplifier back-action is critical, and for axion dark matter searches like ADMX, which require ultra-low-noise, broadband signal routing at cryogenic temperatures. The ability to achieve strong non-reciprocity at such low input powers and moderate bandwidths supports scalable readout chains and noise-resilient detection architectures in both domains \cite{Mallet2009SingleshotQR, PhysRevD.103.032002}.

The spectral response at varying input power levels for the LC-resonator-coupled device is presented in greater detail in Figs.~\ref{fig2} j--l. At an input power of $P=-149\,\mathrm{dBm}$, $S_{32}$ exhibits a Lorentzian-like resonance with the quality factor $Q = 21$, while $S_{31}$ shows a subtle resonance peak (Fig.~\ref{fig2}j).  Compared to the EMP circulator without coupling to an LC resonator, the LC-resonator-coupled configuration achieves significantly higher transmission, with an insertion loss of approximately $6\,\mathrm{dB}$ at the transmission peak (see Appendix~\ref{app:D}). This enhancement can be attributed to the impedance matching between the port impedance and that of the QAH edge state. Transmission can be enhanced to near-unity levels through optimized impedance matching and improved circuit fabrication. 

The broadband non-reciprocal response in the $450-650\,\mathrm{MHz}$ frequency range, observed in both the LC-coupled and uncoupled circulators, arises from interference between the phase-sensitive plasmonic pathway and the parasitic capacitance pathway. For the shorter edge path (port 2 to 3), two LC resonators are capacitively coupled to the EMP resonator, leading to constructive interference with the parasitic transmission. In contrast, for the longer edge path (port 1 to 3), the same coupling occurs; however, the increased path length induces an out-of-phase interaction with the parasitic transmission. This phase mismatch results in destructive interference between the plasmonic and parasitic paths, thereby reducing overall transmission. 

For the LC-coupled-circulator, we find that increasing the input microwave power leads to the gradual emergence of a sharp transmission dip near $543.8\,\mathrm{MHz}$, which reaches its minimum at $P=-119\,\mathrm{dBm}$ (Figs. \ref{fig2} k and l). In addition to the broadband suppression of transmission from port 1 to port 3, this narrow resonance feature produces exceptionally high isolation, exceeding $50\,\mathrm{dB}$ within a bandwidth of approximately $5\,\mathrm{MHz}$. In contrast, this pronounced and narrow transmission feature is absent when the circulator operates without coupling to the LC resonators (Figs.~\ref{fig2} a--f).

\begin{figure}[h]
\centering
\includegraphics[width=0.85\textwidth]{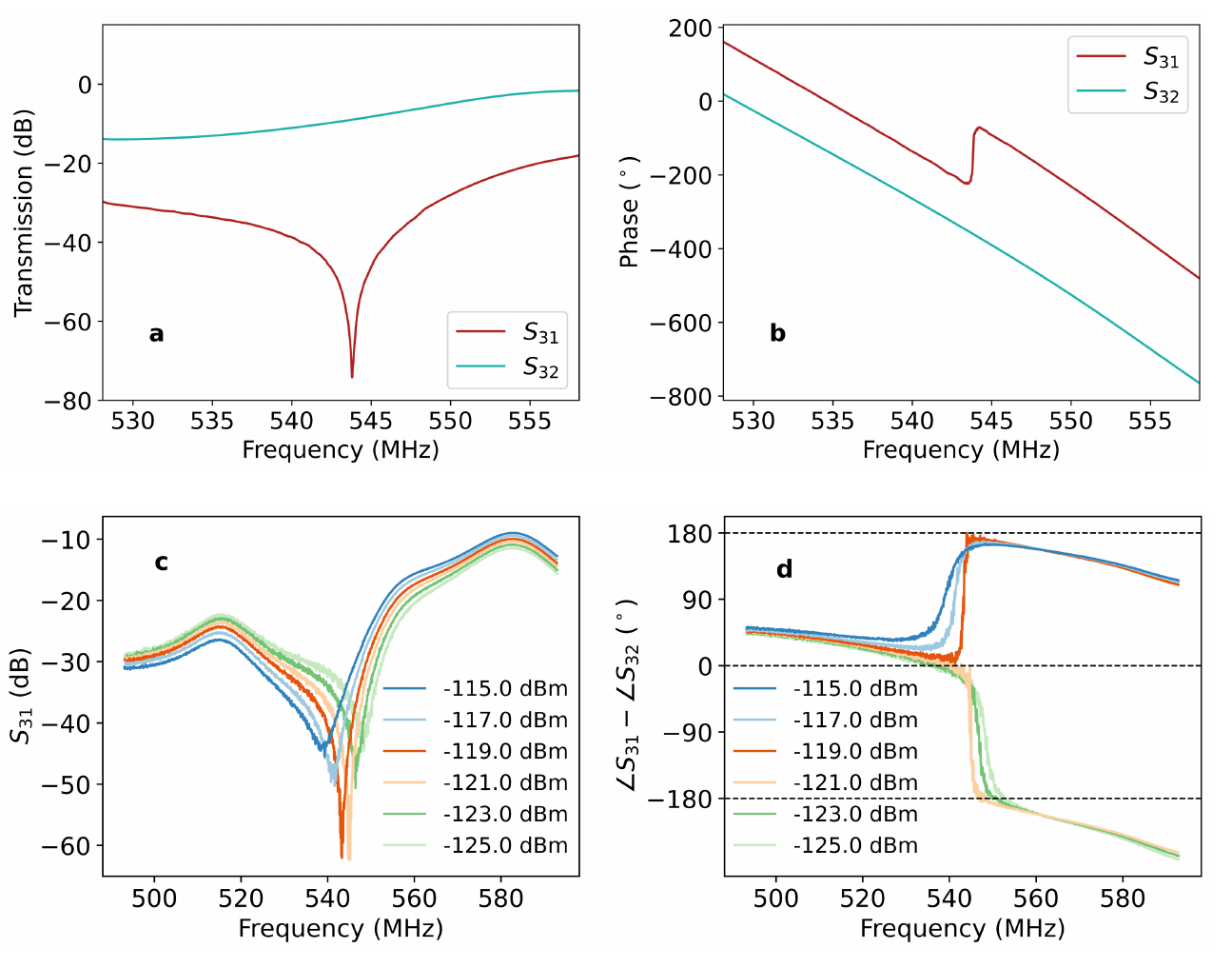}
\caption{{\bf Circulator transmission measurement.} {\bf a, b,} Experimental results for the transmission magnitude and phase for $S_{31}$ and $S_{32}$. The data were taken at an input power of $-120\,\mathrm{dBm}$, corresponding to the estimated RF power at the circulator input ports, where the strongest asymmetric transport behavior is observed. {\bf c, d,} Measured $S_{31}$ and $S_{32}$ magnitudes and their relative phase at input power levels below, near, and above the condition of maximum nonreciprocal transmission.}\label{fig3}
\end{figure}

Next, we further characterize the non-reciprocal response at $f=543.8\,\mathrm{MHz}$ and investigate its physical origin. In particular, we examine how the transmission and phase evolution of $S_{31}$ depend on the input power. Figures~\ref{fig3} a and b show the magnitude and phase of $S_{31}$ and $S_{32}$, respectively, over a narrow frequency range. Figures~\ref{fig3} c and d illustrate the evolution of the $|S_{31}|$ and phase difference $\angle S_{31}-\angle S_{32}$ as a function of input power. A particularly striking observation is that the sharpest transmission contrast appears only within a narrow input-power window (Fig.~\ref{fig3}c). Within this regime, $S_{31}$ undergoes an abrupt phase transition from negative to positive, exhibiting an approximately $180^\circ$ phase shift relative to $S_{32}$ (Fig.~\ref{fig3}d, red curve). This pronounced nonreciprocal transmission, accompanied by a corresponding $\pi$ phase shift, is consistent with strongly asymmetric coupling between the two LC resonators, where the effective coupling becomes highly suppressed in one propagation direction \cite{doi:10.1126/science.aar7709, LiAodong2023Epan, doi:10.1126/sciadv.adr8275}.


We attribute the emergence of the anomalous transmission dip to an effective, power-dependent phase shift in the EMP propagation pathway, which dynamically modulates the coupling between the two LC resonators. As the input microwave power increases, the EMP mode experiences a shift in both velocity and dissipation, altering the accumulated phase along the edge-transport path. The phase and dissipation shifts with increasing power may arise from the presence of charge puddles in proximity to the edge states. The photon-excited charge puddles could slow edge-plasmon propagation and provide dissipation channels through the bulk states \cite{PhysRevB.110.L161403}. Because the EMP-mediated coupling and the parasitic capacitive coupling interfere either constructively or destructively depending on their relative phase, even a small power-induced change in the EMP phase can significantly modify the net interaction strength. At low powers, the plasmonic and parasitic pathways partially cancel, producing moderate suppression of transmission. As the power increases toward ($P \sim -119\ \mathrm{dBm}$), the phase accumulated along the longer EMP path shifts such that the plasmonic signal becomes nearly out of phase with the parasitic contribution. This condition leads to strong destructive interference and substantial suppression of the effective coupling in one propagation direction. The resulting asymmetric transport behavior produces the sharp, unidirectional transmission dip observed in the measurements and is qualitatively consistent with the strongly asymmetric transport regime of the non-Hermitian coupled plasmon-photon system.

\subsection{\label{sec:C}Simulation}
We now turn to simulations of the transmission magnitude and the associated positive phase shift using coupled-mode theory. The system is modeled by the non-Hermitian coupled-mode matrix in Eq.~\eqref{eq1}, which assumes ideal non-reciprocity \cite{Lau2018, YYWang2021, PhysRevLett.134.153801, 2014NatPh..10..394P}. 
Given the almost identical resonance frequencies of the input and output LC resonators (Fig. \ref{figS1}c), we set $\omega_{a} = \omega_{b} = \omega_{0} $ and $\kappa_{a} = \kappa_{b} = \kappa_{0}$. In the EMP resonator, asymmetric coupling emerges due to its chiral nature, leading to differences in propagation time and loss for the $m_1$ and $m_2$ modes (despite $\omega_{m_1}=\omega_{m_2}=\omega_{m}$). These differences depend on the path length and the input microwave power. By applying coupled-mode theory to the steady-state driven response of the non-Hermitian four-mode model in Eq. $ \eqref{eq1} $, and then eliminating the intermediate EMP modes as described in Appendix~$ \ref{app:E} $, the transmission spectrum can be expressed as, 
\begin{equation}
t(\omega) = -\frac{\sqrt{2}\kappa_{0}\alpha_{m} g^2}{\alpha_0^2\alpha_{m}^2-g^4}\label{eq2},
\end{equation}
\smallskip
\noindent
where $\alpha_0 \equiv -i\omega+i\omega_0 + \kappa_0$ characterizes the LC resonator and $\alpha_{m} \equiv -i\omega+i\omega_m + \kappa_m$ characterizes the EMP resonator. Since the longer propagation length of mode $m_1$ is approximately twice that of $m_2$, we adopt the phenomenological approximation $\alpha_{m_1} \approx 2\alpha_{m_2}$ near the operating point of strongest asymmetry. This compactly captures the path-dependent phase accumulation and dissipation, and gives $\alpha_{m} = \sqrt{\alpha_{m_1}\alpha_{m_2}} \approx \sqrt{2}\alpha_{m_2}$.

\begin{figure}[h]
\centering
\includegraphics[width=0.925\textwidth]{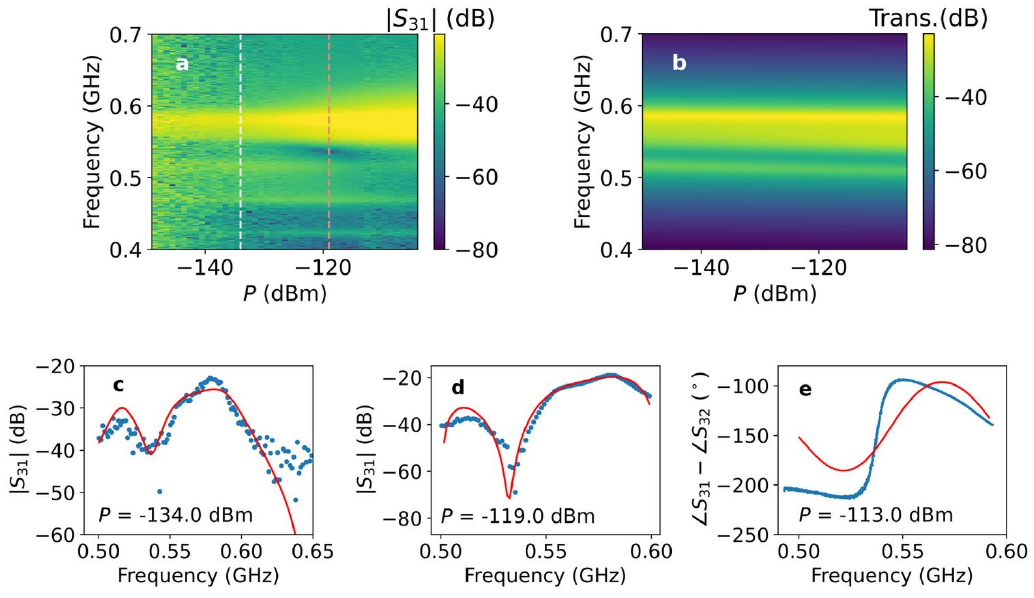}
\caption{{\bf Comparison of data and simulation.} {\bf a, b,} Close-up views of the experimental ({\bf a}) and simulated ({\bf b}) spectral responses as a function of frequency and input power. The white (orange) dashed line indicates the input power corresponding to the measurements shown in {\bf c, d}. {\bf c, d,} Measured $|S_{31}|$ magnitude (blue dots) as functions of frequency at excitation powers of $P = -134$ and $-119\,\mathrm{dBm}$, respectively. The solid red curves represent fits from the theoretical model in Eq. \ref{eq2}. {\bf e,} Measured (blue) phase difference between $S_{31}$ and $S_{32}$ at an excitation power of approximately $P=-113\,\mathrm{dBm}$, and calculated (red) phase response based on Eq. \ref{eq2} using parameters extracted from the fit presented in {\bf d}. See text for details.}\label{fig4}
\end{figure}

We find that the measured transmission spectra are qualitatively in agreement with the transmission response derived from the effective two-site Hatano-Nelson-type model, indicating that the circulation is consistent with a mechanism dominated by asymmetric coupling between the two LC resonators mediated by the EMP mode (Figs. \ref{fig4} a and b). At low excitation power, the measured transmission closely matches the model, with fitted parameters: $g/2\pi=5.3\,\mathrm{MHz}$, $\kappa_0/2\pi=3.0\,\mathrm{MHz}$, and $\kappa_m/2\pi=1.0\,\mathrm{MHz}$ (Fig. \ref{fig4}c). As the input power increases, the fitted coupling strength $g$ rises to $6.9\,\mathrm{MHz}$, while $\kappa_0/2\pi$ remains at $3.0\,\mathrm{MHz}$ and $\kappa_m/2\pi$ increases to $3.3\,\mathrm{MHz}$, respectively, at the point of maximum isolation (Fig. \ref{fig4}d). 

This effective model also captures the phase behavior observed in the experimental data, particularly in the regime of strongest asymmetric transport, as shown in Fig. \ref{fig4}e. It is worth noting that the phase data in Figs. \ref{fig4}, a and b, are not sampled densely enough to enable a reliable comparison between experiment and simulation. To address this, we took a high-resolution data scan and plotted the phase response of $S_{31}$ after subtracting the phase of $S_{32}$, thereby removing the line delay. This allows for a direct comparison with the theoretical result. The residual mismatch between experiment and theory at higher powers is likely due to power-dependent plasmon damping and coupling effects that lie outside the minimal model.

\section{Conclusion}
In this work, we demonstrated a novel circulation mechanism in a topological circulator comprising a QAH edge magnetoplasmonic resonator coupled to LC resonators. Within an input power range of $-150$ to $-120\,\mathrm{dBm}$, the circulator maintains a relatively stable bandwidth, passband center frequency, and isolation, indicating robust performance under low-power operating conditions. This power independence suggests that the device is well-suited for integration into  superconducting quantum information science platforms and high-energy physics experiments. If fully harnessed, this mechanism could facilitate the realization of a topological circulator with near-unitary forward transmission and exceptionally high isolation in the reverse direction. 

In the present device geometry, the operating frequency is primarily determined by the edge magnetoplasmon velocity and the dimensions of the topological insulator mesa. The demonstrated operation near 0.5 GHz serves as a proof of principle for realizing effective non-Hermitian coupling and nonreciprocal microwave transport using magnetic topological insulators. Furthermore, we note that compact on-chip nonreciprocal devices operating in the sub-GHz frequency range may find applications in superconducting quantum circuits that employ low-frequency signals for flux control, biasing, and tunable coupling. In particular, the frequency range demonstrated here overlaps with that of fluxonium-based architectures, where control-line-limited relaxation and flux-controlled interactions have recently attracted significant interest \cite{PhysRevX.9.041041,PRXQuantum.5.040342,PRXQuantum.5.020326,azar2026characterizationcomparisonenergyrelaxation}. Looking forward, the operating frequency can be extended to higher microwave frequencies through device scaling and QAH material engineering of the edge plasmon properties. 

In conclusion, our results highlight the significant potential of chiral plasmons to provide enhanced control over signal transmission and isolation across a wide dynamic range.

\begin{acknowledgments}
We would like to thank Yaniv Rosen, Kristin Beck, and Enrico Rossi for helpful discussions. This work was performed under the auspices of the US Department of Energy by Lawrence Livermore National Laboratory under Contract No. DE-AC52-07NA27344. The project was supported by the Laboratory Directed Research and Development (LDRD) programs of LLNL (21-ERD-014 and 26-ERD-010) and HEP Laboratory Detector Research and Development program. The authors also acknowledge support from NSF Convergence Accelerator program under Grant 2040737.
\end{acknowledgments}

\appendix

\section{\label{app:A}Experimental configuration and device fabrication}

\begin{figure}[h]
\centering
\includegraphics[width=1\textwidth]{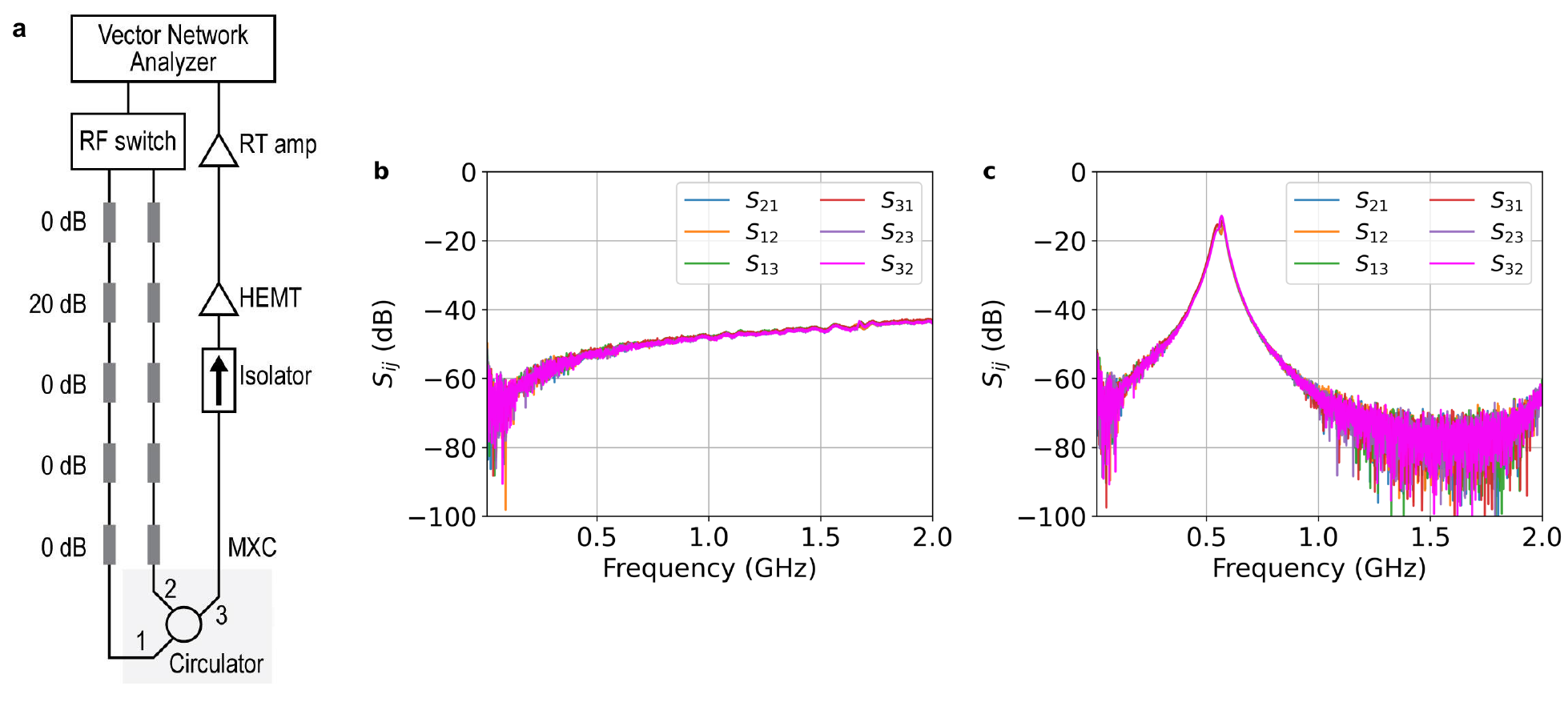}
\caption{{\bf Experimental configuration and the transmission spectra of the device without and with the inductors measured at room temperature.} {\bf a,} Experimental setup showing the circulator mounted on the mixing-chamber plate of a dilution refrigerator, which cools the device to below 10 mK. {\bf b,} Complete six-way transmission responses of the $100$-$\mu m$-radius circulator without inductors measured at room temperature. {\bf c,} The transmission spectra $S_{ij}$ for the $100$-$\mu \mathrm{m}$-radius circulator with inductors, showing a frequency response resulting from the LC impedance-matching circuits. }\label{figS1}
\end{figure}

The transmission of the circulator is probed by measuring the $S_{21}$ transmission parameters for various powers. The experimental configuration is plotted in Fig. \ref{figS1}a. The $S_{21}$ measurements were performed on two ports of the circulator with an RF switch located at room temperature. The input microwave signals are transmitted through coaxial cables with a $20\,\mathrm{dB}$ attenuator at room temperature (not shown in Fig. \ref{figS1}a) and a $20\,\mathrm{dB}$ attenuator at $4\,\mathrm{K}$. The total attenuation from the vector network analyzer down to the sample is about $-64\,\mathrm{dB}$. The power values expressed in decibels in this letter represent the estimated power at the input port of the circulator. The circulator output goes through a cryogenic isolator at the still stage and is amplified by a HEMT amplifier (LNF-LNC0.3-14A, Low Noise Factory) with a noise temperature of $\sim 2\,\mathrm{K}$ at the $4\,\mathrm{K}$ stage. At room temperature, the transmission signal is further amplified by a low-noise room temperature amplifier with a total gain of $20\,\mathrm{dB}$ (ZX60-4016E-S+, Mini-Circuits).

Topological insulators are materials with an insulating bulk and topologically protected conducting surface or edge states arising from strong spin–orbit coupling and a nontrivial electronic band topology \cite{Qu2010, PhysRevLett.121.037001, PhysRevLett.109.246602}. Magnetic topological insulators are topological insulators in which magnetic order breaks time-reversal symmetry, leading to gapped surface states and enabling exotic phenomena such as the QAH effect. The 6 quintuple layers of magnetic topological insulator Cr$_{0.12}$(Bi$_{0.26}$Sb$_{0.62}$)$_2$Te$_3$ thin films were grown using ultrahigh vacuum molecular beam epitaxy. Epi-ready semi-insulating ($\rho > 10^6\,\Omega\,\mathrm{cm}$) GaAs (111) B substrates were pre-annealed in the chamber at up to $600 \ ^\circ \rm{C}$ under Te-rich environment. During the co-evaporated growth, high-purity Bi ($99.9999\%$), Te ($99.9999\%$), Cr ($99.99\%$) were evaporated from Knudsen effusion cells while Sb ($99.999\%$) was evaporated from a cracker cell. 
Upon material growth, the circulator devices were fabricated via standard photolithography. The mesa disk was first patterned by photolithography and then etched by CF$_4$/Ar using reactive ion etching. The coplanar waveguides and ground planes were patterned by photolithography, and $10\,\mathrm{nm}$ of Ti and $100\,\mathrm{nm}$ of Au were deposited using electron beam deposition, followed by a liftoff process. The Hall characterization of the same batch of magnetic topological insulator wafer can be found in reference \cite{PhysRevLett.128.217704}. All measurements presented in this work were performed under an external magnetic field of approximately 40 mT, generated by attaching a permanent magnet plate underneath the QAH mesa. 

\section{\label{app:B}Room temperature device characterization}
\begin{figure}[h]
\centering
\includegraphics[width=0.65\textwidth]{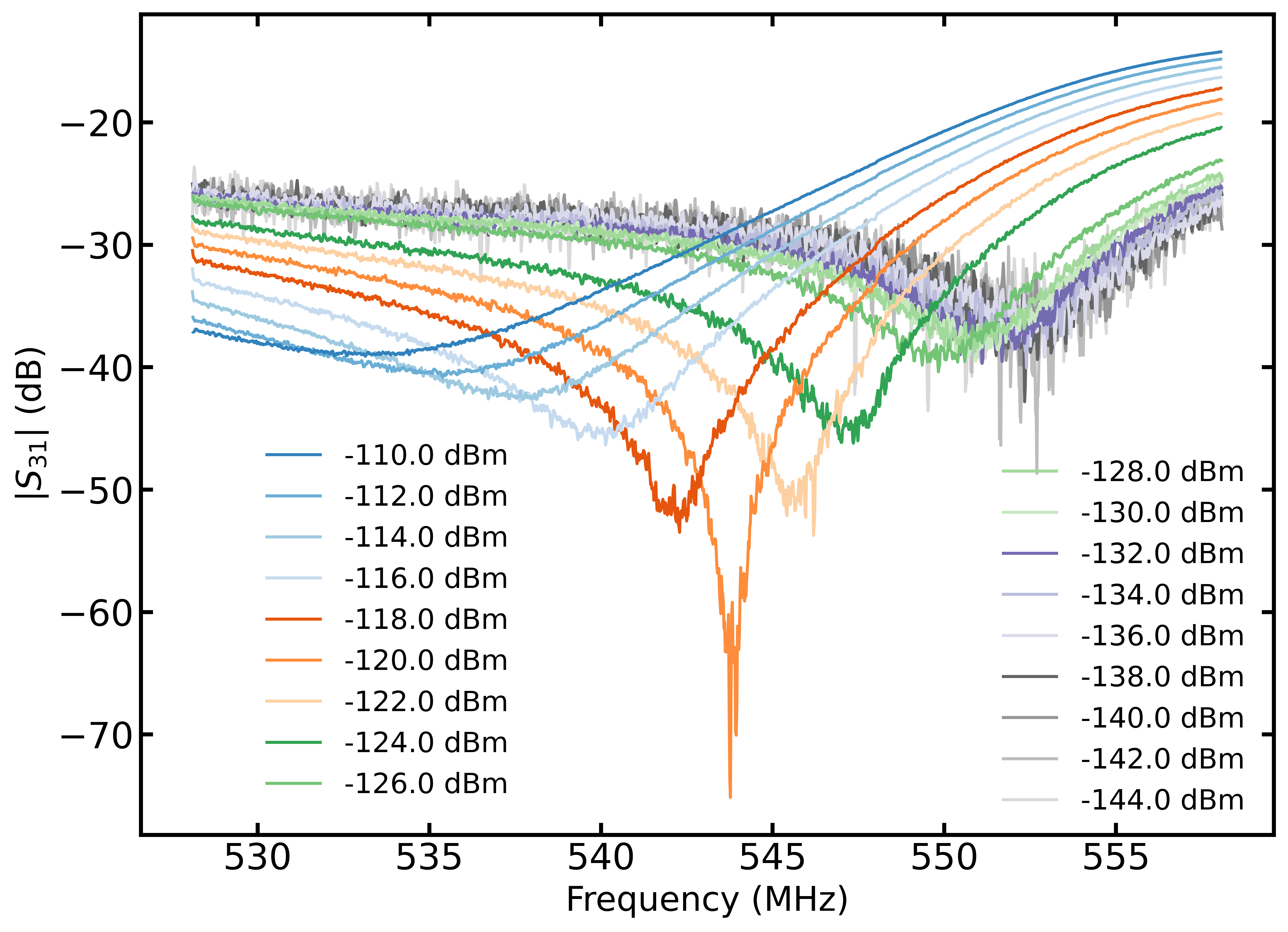}
\caption{{\bf Transmission spectra from port 1 to 3 at various input power levels.} The transmission dip shows a power-dependent frequency shift.}\label{figS2}
\end{figure}
\smallskip

\noindent
The circulator sample was initially characterized without a matching inductor and its impedance was expected to be on the order of the Hall resistance, $\rho_{xy}= 25\,\mathrm{k}\Omega$. Since the impedance of the measurement circuit was $50\,\Omega$, the lack of a matching circuit at the input results in approximately $99\%$ of the power being reflected at the input port of the circulator. As a control, we first measured all microwave $S$ parameters before mounting the inductors, observing that all ports are equivalent with a transmission magnitude approximately $-50\,\mathrm{dB}$ at $530\,\mathrm{MHz}$, as shown in Fig. \ref{figS1}b.

\begin{figure}[h]
\centering
\includegraphics[width=0.85\textwidth]{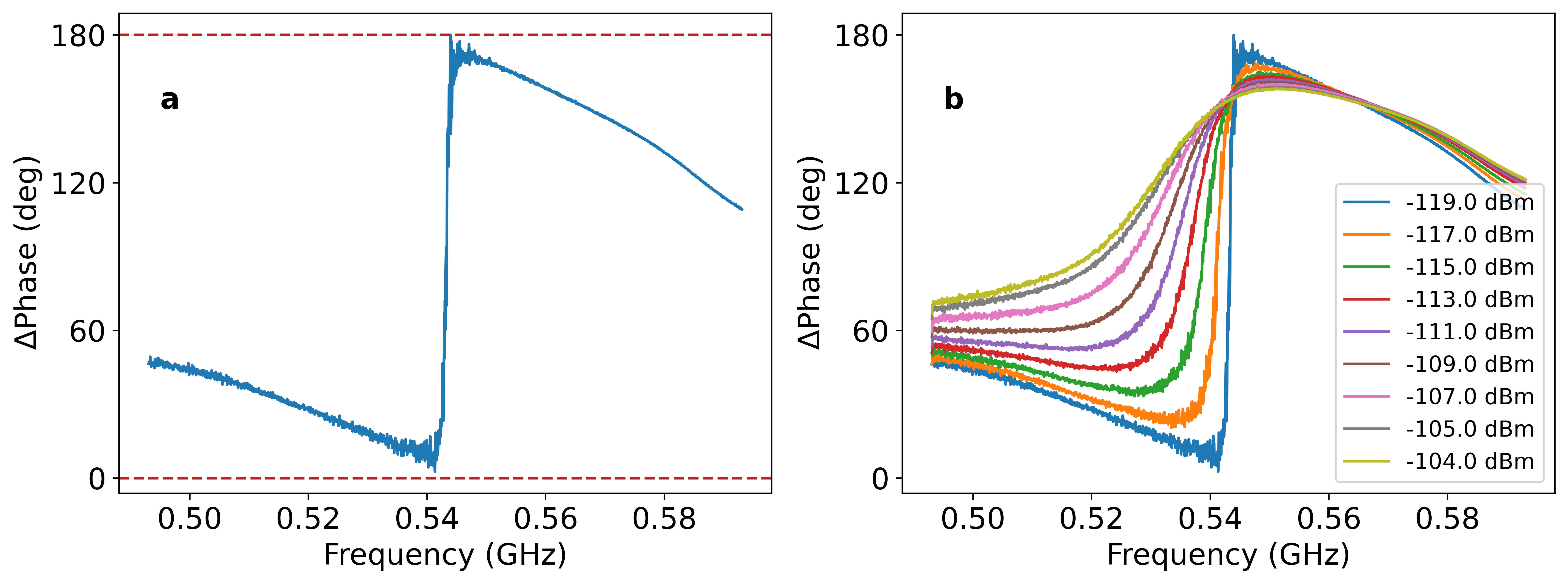}
\caption{{\bf The measured phase difference between $S_{31}$ and $S_{32}$ at various input power levels.}  {\bf a,} $\Delta \mathrm{Phase} = \angle S_{31} - \angle S_{32}$ at an input power of $P = -119\,\mathrm{dBm}$. {\bf b,} The phase difference at various input power levels above the EP condition. }\label{figS3}
\end{figure}
\smallskip

We then implemented a matching network at the input ports of the circulator by utilizing $3$ lumped element $180\,\mathrm{nH}$ inductors in series (Fig. \ref{fig1}a). The inductors coupled with the effective capacitance of the circuit serve as an impedance matching network for the $50\,\Omega$ coaxial lines and the high impedance of the circulator. Physically, the
matching process utilizes the resonance created by the inductors and the effective stray capacitance of the circuit. The resonance is tuned to coincide with the circulator's edge magnetoplasmonic resonance frequency, effectively increasing the energy
coupled into the circulator. Increasing the energy coupled into the circuit increases the signal-to-noise ratio. In practice, better
coupling improves the insertion loss associated with reflections arising from impedance mismatch. The appropriate inductor value was determined by testing different inductors on a similar sample and collecting transmission measurements at room temperature. Figures \ref{figS1}, b and c present the six-way transmission spectra measured without and with the inductors, respectively. At room temperature, the LC network exhibits a resonant frequency of $f = 566\,\mathrm{MHz}$ and a quality factor of $Q = 21$. 


\section{\label{app:C}Power dependent transmission measurements for $S_{31}$}
\noindent
Figure \ref{figS2} shows the magnitude of $S_{31}$ plotted as a function of frequency for various input power levels at temperatures below 10 mK. When the signal is applied at port 1, both the center frequency and bandwidth exhibit a clear dependence on excitation power. In contrast, transmission from port 2 shows no significant power dependence.

The phase difference between $S_{31}$ and $S_{32}$ at different excitation powers is illustrated in Fig. \ref{figS3}. At the maximum $S_{31}$ transmission dip ($P = -119\,\mathrm{dBm}$), a phase shift of approximately $180^{\circ}$ is observed (Fig. \ref{figS3}a). The slope of phase change decreases when $P$ exceeds $-119\,\mathrm{dBm}$ (Fig. \ref{figS3}b).
\begin{figure}[h]
\centering
\includegraphics[width=0.6\textwidth]{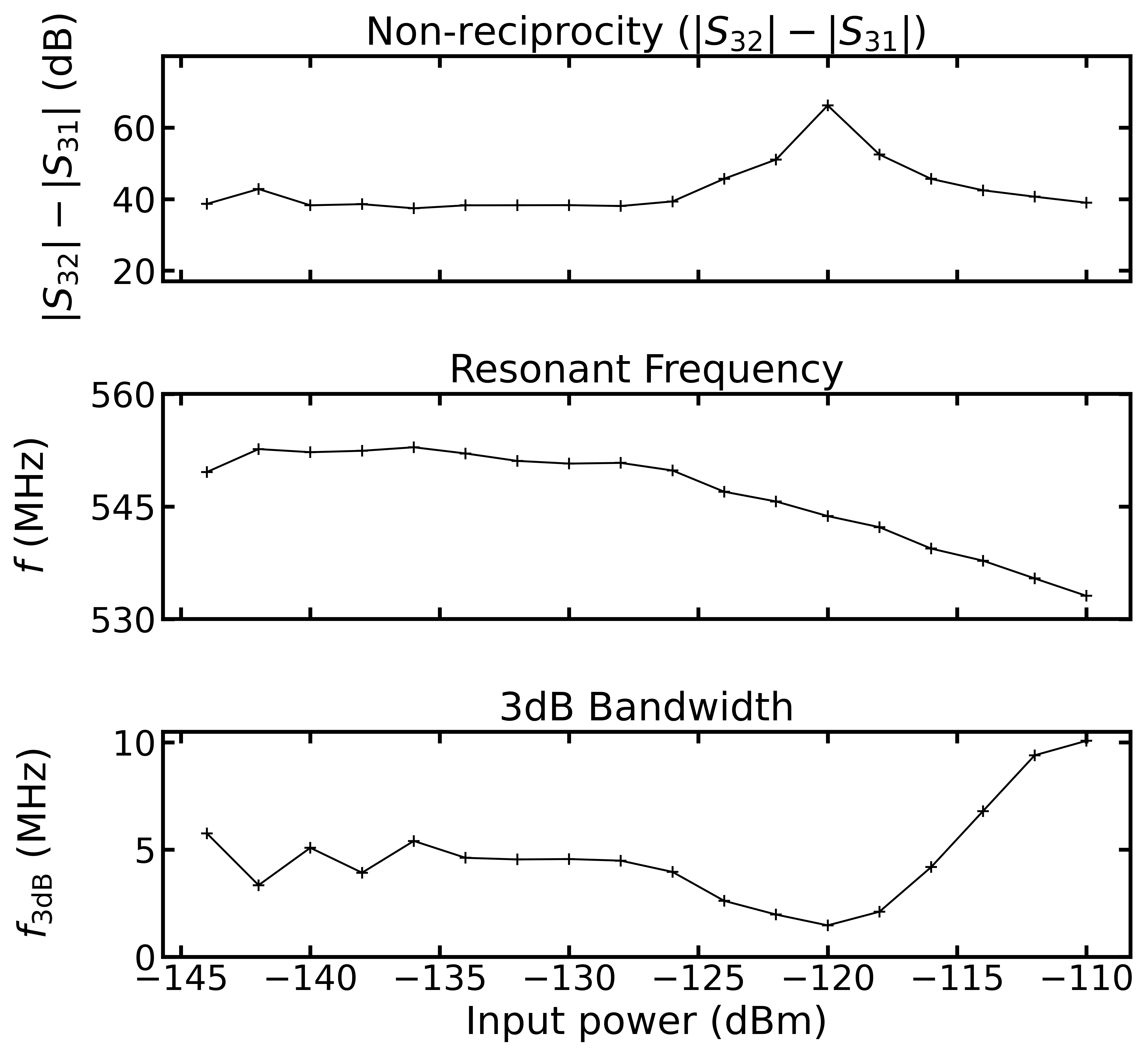}
\caption{{\bf Parameters extracted from the transmission spectra as a function of input power.} The non-reciprocity, defined as the difference between $|S_{32}|$ and $|S_{31}|$, along with the frequency dip in $|S_{31}|$ and the $3\,\mathrm{dB}$ bandwidth, are plotted versus input power.}\label{figS4}
\end{figure}

Figure \ref{figS4} presents the non-reciprocity $|S_{32}|-|S_{31}|$, the resonant frequency shift, and the $3\,\mathrm{dB}$ bandwidth as functions of input power. Notably, within the input power range of $-145$ to $-125\,\mathrm{dBm}$, both the circulator’s bandwidth and center frequency remain relatively stable, indicating a power-independent regime. This suggests that the circulator can operate reliably within power levels relevant to superconducting quantum information systems and high-energy physics experiments.

\section{\label{app:D}Insertion loss characterization}

\begin{figure}[h]
    \centering
    \includegraphics[width=0.6\linewidth]{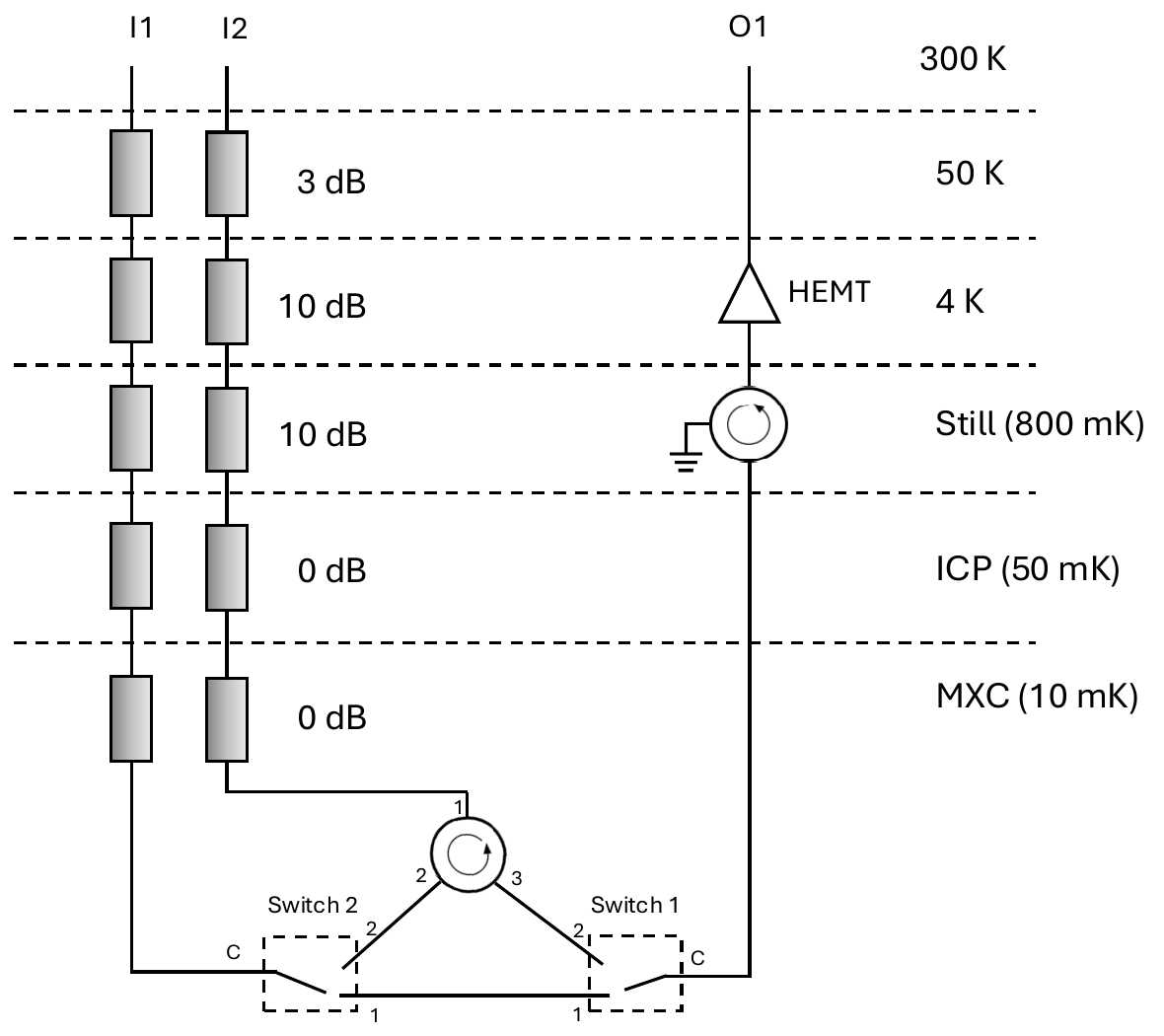}
    \caption{{\bf Experimental setup for insertion loss measurements.} The circulator is cooled to below $30\,\mathrm{mK}$. A network analyzer is used to measure transmission through the receiver and two RF switches enable bypassing the circulator. Measurement of transmission with and without the circulator enable a direct measurement of the insertion loss of the circulator.}
    \label{fig:receiver_diagram}
\end{figure}

\begin{figure}[h]
    \centering
    \includegraphics[width=0.8\linewidth]{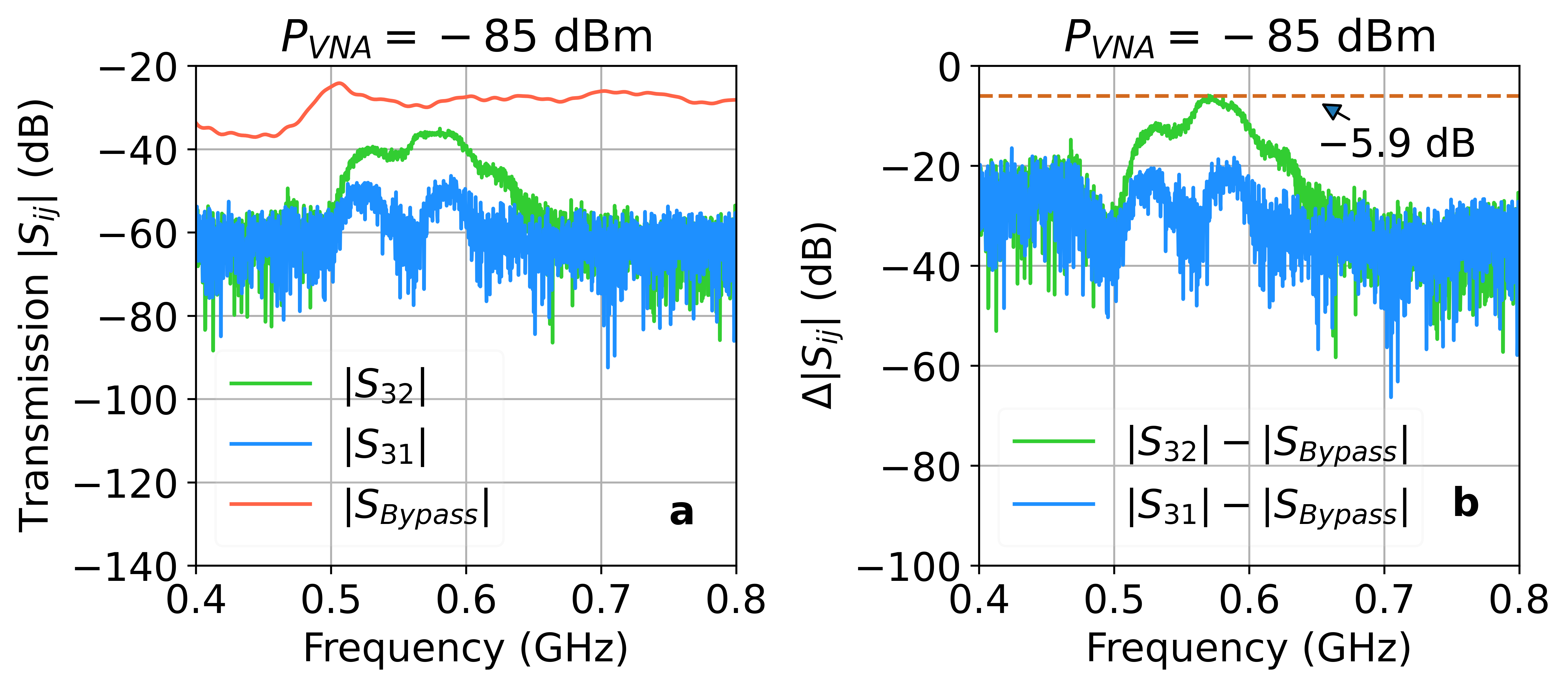}
    \caption{{\bf Insertion loss of the circulator at approximately 10 mK.} {\bf a,} Transmission of the circulator and the short bypass cable at the VNA power of $P_{VNA} = -85\,\mathrm{dBm}$. {\bf b,} An insertion loss of $5.9\,\mathrm{dB}$ is observed at the transmission peak of $|S_{32}|$ with an isolation of $\sim 20\,\mathrm{dB}$.}
    \label{fig:circ_loss}
\end{figure}

\noindent
The insertion loss of the circulator was measured in a different Bluefors dilution refrigerator after the sample, with the LC resonators attached, had been stored for more than two years at ambient temperature in a low-vacuum desiccator. The characterization diagram is shown in Fig~\ref{fig:receiver_diagram}. The circulator was cooled to below $30\,\mathrm{mK}$. A network analyzer was used to perform $S_{21}$ transmission measurements through the system. To determine the insertion loss of the circulator, a transmission measurement was made through the transmission port ($I_1$) and the proportion of power transmitted through the receiver was measured at the output ($O_1$). Two cryogenic switches (Radial R585433210) enabled swapping between the circulator ports and a short cable. In addition, a second transmission line ($I_2$) was used to measure the isolation of the circulator. On the output line ($O_1$), the signal was amplified by a HEMT amplifier mounted to the 4 K stage, followed by a room temperature amplifier. An isolator before the HEMT amplifier was used to reduce standing waves within the receiver. Broadband transmission measurements were made of both the $S_{32}$ and $S_{31}$ of the circulator. At approximately $10\,\mathrm{mK}$, we observe a $\sim 2\,\mathrm{dB}$ difference in transmission between input lines $I_1$ and $I_2$.

To determine the insertion loss of the circulator, transmission measurements were performed with the circulator both in the receiver circuit and bypassed, while applying a range of input powers. Any additional transmission losses observed when the circulator was switched in were attributed to its insertion loss. As shown in Fig.~\ref{fig:circ_loss}, 
the insertion loss of the circulator was found to be approximately $5.9 \pm 2\,\mathrm{dB}$ with an isolation of $\sim 20\,\mathrm{dB}$ in the low power regime. 

\section{\label{app:E} Effective Hatano-Nelson description and derivation of the transmission model}

In this Appendix, we show how the four-mode model in Eq.~(\ref{eq1}) reduces to an effective two-resonator description with asymmetric directional couplings, and how the same reduction yields the transmission formula in Eq.~(\ref{eq2}). Equation~(\ref{eq1}) is a minimal device-level coupled-mode model for the two LC resonators and the two directional edge-magnetoplasmon (EMP) propagation channels.

We begin from the non-Hermitian four-mode matrix in the basis $\Psi=\left(a,m_1,b,m_2\right)^T$,
\begin{equation}
H_{\rm eff}/\hbar=
\left(
\begin{array}{cccc}
\omega_a-i\kappa_a & g & 0 & 0 \\
0 & \omega_{m_1}-i\kappa_{m_1} & g & 0 \\
0 & 0 & \omega_b-i\kappa_b & g \\
g & 0 & 0 & \omega_{m_2}-i\kappa_{m_2}
\end{array}
\right),
\label{eq:App_H4}
\end{equation}
where $a$ and $b$ denote the two LC resonators, $m_1$ and $m_2$ denote the two directional EMP channels, and $g$ is the LC-EMP coupling rate. In the experiment, the resonators are nearly identical, so we take $\omega_a=\omega_b=\omega_0$ and $\kappa_a=\kappa_b=\kappa_0$. The off-diagonal structure of Eq.~(\ref{eq:App_H4}) describes the directional sequence
\begin{equation}
a \rightarrow m_1 \rightarrow b \rightarrow m_2 \rightarrow a.
\end{equation}
The two EMP channels correspond to inequivalent chiral propagation paths around the QAH mesa, and therefore generally acquire different phase shifts and dissipation.

For a probe frequency $\omega$, we define
\begin{equation}
\alpha_{j}(\omega)\equiv -i\omega+i\omega_{j}+\kappa_{j},
\qquad
j\in\{a,b,m_1,m_2\}.
\end{equation}
It follows that the steady-state coupled-mode equations for a coherent drive, $s_{in}$, incident on resonator $a$ are
\begin{align}
\alpha_a a + ig m_1 &= s_{\rm in},
\label{eq:App_a}
\\
\alpha_{m_1} m_1 + ig b &= 0,
\label{eq:App_m1}
\\
\alpha_b b + ig m_2 &= 0,
\label{eq:App_b}
\\
ig a + \alpha_{m_2} m_2 &= 0.
\label{eq:App_m2}
\end{align}
Eliminating the EMP amplitudes ($m_1, m_2$) gives
\begin{align}
\alpha_a a + \frac{g^2}{\alpha_{m_1}}\,b &= s_{\rm in},
\label{eq:App_red1}
\\
\alpha_b b + \frac{g^2}{\alpha_{m_2}}\,a &= 0.
\label{eq:App_red2}
\end{align}
The two resonators are therefore coupled by the effective directional couplings
\begin{equation}
\lambda_{ab}^{\rm eff}(\omega)=\frac{g^2}{\alpha_{m_1}(\omega)},
\qquad
\lambda_{ba}^{\rm eff}(\omega)=\frac{g^2}{\alpha_{m_2}(\omega)},
\label{eq:App_lambda}
\end{equation}
and because the circulation paths have different lengths and dissipation, $\alpha_{m_1}\neq\alpha_{m_2}$, leading to asymmetric coupling. Hence, Eq.~(\ref{eq1}) reduces to a two-site Hatano-Nelson-type model with effective matrix
\begin{equation}
M_{\rm eff}(\omega)=
\left(
\begin{array}{cc}
\alpha_a(\omega) & \lambda_{ab}^{\rm eff}(\omega) \\
\lambda_{ba}^{\rm eff}(\omega) & \alpha_b(\omega)
\end{array}
\right).
\label{eq:App_H2}
\end{equation}

The effective asymmetric coupling between the two resonators can be interpreted as a two-site reduction of the canonical Hatano-Nelson Hamiltonian,
\begin{equation}
 H/\hbar= \sum_{n}\left(\omega_0 \hat{a}_n^{\dagger}\hat{a}_n+\lambda_{ab}\hat{a}_{n+1}^{\dagger}\hat{a}_n+\lambda_{ba}\hat{a}_{n}^{\dagger}\hat{a}_{n+1}\right),
  \label{eqB1}
\end{equation}
\smallskip
\noindent where $\omega_0$ is the resonant frequency of the $n$th resonator; $\hat{a}_n^{\dagger}$ and $\hat{a}_n$ are the bosonic creation and annihilation operators at the $n$th resonator; $\lambda_{ab}$ and $\lambda_{ba}$ represent non-Hermitian coupling between adjacent resonators. In our system, the effective couplings are set by the EMP propagation paths and are generally unequal, $\lambda_{ab} \neq \lambda_{ba}$. Thus, the device realizes the minimal two-site analogue of Hatano-Nelson physics in an open, driven, and dissipative system. 

The same reduction also yields the transmission formula used to fit the data. Solving Eqs.~(\ref{eq:App_red1}) and (\ref{eq:App_red2}) gives
\begin{equation}
\frac{b}{s_{\rm in}}
=
-\frac{g^2\alpha_{m_1}}
{\alpha_a\alpha_b\alpha_{m_1}\alpha_{m_2}-g^4}.
\label{eq:App_ratio_general}
\end{equation}
For nearly identical LC resonators, $\alpha_a=\alpha_b\equiv\alpha_0$, where
\begin{equation}
\alpha_0=-i\omega+i\omega_0+\kappa_0,
\end{equation}
so that
\begin{equation}
\frac{b}{s_{\rm in}}
=
-\frac{g^2\alpha_{m_1}}
{\alpha_0^2\alpha_{m_1}\alpha_{m_2}-g^4}.
\label{eq:App_ratio}
\end{equation}
Introducing $\alpha_m$ through
\begin{equation}
\alpha_m^2\equiv\alpha_{m_1}\alpha_{m_2},
\label{eq:App_alpham}
\end{equation}
and using $\alpha_{m_1}=2\alpha_{m_2}$ near the operating point (see main text), we obtain Eq.~(\ref{eq2});
\begin{equation}
t(\omega)=
-\frac{\sqrt{2}\kappa_0\alpha_m g^2}
{\alpha_0^2\alpha_m^2-g^4},
\end{equation}
where the factor $\kappa_0$ comes from the canonical input--output transmission amplitude normalization for a resonator coupled to a transmission line. 



\bibliographystyle{unsrt}

\end{document}